# A Systematic Review of Empirical Research on Graphing Numerical Data in K-12 STEM Education

Verena Ruf[ai], Dominik Thüs[bj], Sarah Malone[cj], Stefan Küchemann[di], Sebastian Becker-Genschow[e], Markus Vogel[f], Roland Brünken[gj], Jochen Kuhn[hi]

[a] Corresponding author. Email: v.ruf@physik.uni-muenchen.de, LMU, Edmund-Rumpler-Straße 13, 80939 Munich, Germany

[b] Email: dominik.thues@uni-saarland.de

[c] Email: s.malone@mx.uni-saarland.de

[d] Email: s.kuechemann@physik.uni-muenchen.de

[e] Digital Education Research, Faculty of Mathematics and Natural Sciences, University of Cologne, Herbert-Lewin-Straße 10, 50931 Cologne, Germany. Email: sebastian.becker-genschow@uni-koeln.de

[f] Fakultät für Natur- und Geisteswissenschaften, Pädagogische Hochschule Heidelberg, Keplerstraße 87, 69120 Heidelberg, Germany. Email: vogel@ph-heidelberg.de

[g] Email: r.bruenken@mx.uni-saarland.de

[h] Email: jochen.kuhn@physik.uni-muenchen.de

[i] Faculty of Physics / Chair of Physics Education Research, LMU Munich, Munich, Edmund-Rumpler-Straße 13, 80939 Munich, Germany.

[j] Department of Education, Saarland University, Campus Saarbrücken, Building A4 2, 66123 Saarbrücken, Germany.

**Abstract**

Graphs are essential representations in the professions and education concerning the science, technology, engineering, and mathematics (STEM) disciplines. Beyond their academic relevance, graphs find extensive utility in everyday scenarios, ranging from news media to educational materials. This underscores the importance of people's being able to understand graphs. However, the ability to understand graphs is connected to the ability to create graphs. Therefore, in school education, particularly in STEM subjects, not only the understanding but also the skill of constructing graphs from numerical data is emphasized. Although constructing graphs is a skill that most people do not require in their everyday lives and professions, it is a well-established student activity that has been empirically studied several times. Therefore, since a synthesis of the research findings on this topic has not yet been conducted, a summary of the studies investigating graphing via various viewpoints and differing methods could be a valuable contribution. To provide an overview of the empirical literature on this important topic, our systematic review identifies how the construction of convention-based graphical representations of numerical data, referred to as graphing, has been studied in previous research, how effective graphing is, and which types of difficulties are encountered by students. Based on these aspects, we defined inclusion criteria that led to 50 peer-reviewed empirical studies on graphing in K–12 STEM education found in SCOPUS, ERIC, and PsychInfo. Graphing instruction seemed to be beneficial for student learning, not only improving graph construction but also graph interpretation skills. However, the students experienced various difficulties during graphing, both during graph construction and the interpretation and usage of data.

# Introduction

General mathematical skills, such as the competent handling of numerical data and their external representations (Chalkiadaki, 2018), are key skills in the 21st century for thinking critically and using information adequately (Program for International Student Assessment, 2021). Numerical data are often represented graphically to provide a comprehensive and easily accessible overview of a topic. Unlike informal representations, such as sketches, formal graphical data representations follow certain conventions, and to extract information correctly, it is necessary to be familiar with these conventions. We define *graphing* as the construction of convention-based graphical representations of numerical data, a key aspect of graphing competence (Glazer, 2011). *Graphing competence* contains both the interpretation and creation of convention-based graphical representations (Glazer, 2011) and is an important skill in K–12 education, especially in science, technology, engineering, and mathematics (STEM) disciplines. Consequently, both activities should be practiced extensively in class (Díaz-Levicoy et al., 2018; Dossey et al., 2016; Glazer, 2011; Kultusministerkonferenz, 2012). In addition, the ability to interpret numerical data is fundamental to understanding everyday statistical phenomena, such as stock markets, the progression of a pandemic, and risk evaluation (Schüller et al., 2019). Therefore, numerical data are highly relevant beyond STEM fields, and the ability to understand data is an essential part of everyday life (Friedrich et al., 2024). Several theoretical scientific works and multiple empirical studies have examined this graphing as a student activity in K–12 education. However, empirical research on this topic is diverse using different study designs and analysis methods. In this systematic review, we aim to synthesize existing empirical research on the use of student graphing of numerical data in K–12 STEM education and its effectiveness as an educational method by reviewing how this skill has been employed in various contexts, what its effectiveness has been in terms of different learning outcomes, and which difficulties students and teachers have reported.

**Representations in STEM Education**

Representations are common in STEM education (Rau, 2017) and serve multiple functions, such as memory support, facilitating inferences, or making discoveries (Tversky, 1997). Learning material can, for example, consist of text, equations, and graphical representations – such as pictorial or graphical representations of numerical data. Therefore, such learning material can be considered in the context of learning with multiple representations.

### *Multiple Representations in Education*

Two prominent theoretical models that deal with multiple representations in learning are the *Integrated Model of Text and Picture Comprehension* (ITPC; Schnotz, 2005) and the *Cognitive Theory of Multimedia Learning* (CTML; Mayer, 2005, 2014). Both models, distinguish between symbolic and analogue representations (usually text and pictures) and predict the learning benefits of the simultaneous use of both types of representations owing to the construction of dual mental representations, which enhances the storage and retrieval of information (the multimedia effect; Mayer, 2005). Ott et al. (2018) showed that the multimedia effect also accounts for problem-solving using text and symbolic mathematical representations. Different types of representations accentuate different information (Zhang & Norman, 1994) thus activating different cognitive functions (Zhang, 1997). Understanding this thinking process is also important when considering graphical representations of data in STEM education (Duval, 2006). For example, data is interpreted differently depending on whether it is presented in a bar or a line graph (Shah & Freedman, 2011).

Besides their representational advantages, multiple representations can fulfil various other pedagogical functions. Ainsworth (1999) defined three different functions that provided or self-generated multiple representations can have during learning. First, two (or more) representations can complement each other. In this regard, Ainsworth (1999) distinguished between representations that involve complementary processes and those that contain complementary information. In complementary processes, diverse representations may help perform different tasks, support unique learners' characteristics, or accommodate specific strategies that learners may want to use. Multiple representations with complementary information are not redundant; rather, each individual representation provides unique essential information for learning. Second, multiple representations can constrain one another's interpretations to prevent misunderstandings—for example, by adding a familiar type of representation to the introduction of a new one. Moreover, representations constrain one another via their properties—for example, if one representation is more specific than another (Ainsworth, 1999). Third, representations can be used to deepen understanding via abstraction, extension, or relations. Abstraction, as a function of representations, entails the reorganization of information. If the reason for using multiple representations is extension, then knowledge is enhanced—for example, by being transferred to another context. Relating information between representations means translating between representations. Finally, representations can be used for more than one function simultaneously (Ainsworth, 2006).

These functions are also a part of Ainsworth's (2006) design, function, and task (DeFT) framework. This framework proposes that aspects of design, functions, and tasks should influence the creation of learning material that includes multiple representations. Apart from the aforementioned functions, DeFT considers the following five design parameters: the number of representations, how the information is distributed across representations, the types of representations, the order of representations, and the amount of support needed to translate between representations (Ainsworth, 2006). The cognitive tasks that learners are presented with when they encounter a (novel) representation include understanding how information is represented, how this information relates to the corresponding domain, how to choose a suitable representation, and how to construct a representation (Ainsworth, 2006). Self-generated graphs have the potential to serve all the functions from Ainsworth's approach, depending on the characteristics of the visualized information, such as design, the relation to other given representations, and learner characteristics. Moreover, graphing might include all cognitive tasks suggested by Ainsworth's framework (2006), as learners have to first choose and then correctly generate a graphical representation, which requires knowledge of how to correctly present information and relate the information to the domain as well as to other representations. Graphing thus leads to learners handling multiple external representations, which can promote learning in different ways.

Students need certain competencies to effectively learn with multiple representations, such as the ability to describe and use (multiple) representations appropriately, which involves the ability to choose a representation based on the context, task demands, as well as on personal ability and goals (Rau, 2017). This is an important instructional goal for students (diSessa, 2004), especially due to its relevance for scientific education (diSessa & Bruce, 2000) and practice (Gooding, 2004). The ability to use multiple representations is an essential skill for communicating information (de Vries & Masclet, 2013) and solving problems (Duval, 2006; Zhang, 1997). For example, sixth-grade students were more likely to solve math problems when they used schematic instead of pictorial representations (Hegarty & Kozhevnikov, 1999). Difficulties in dealing with representations might be due to lacking metarepresentational competence (diSessa, 2004). For example, Duval (2006) distinguishes between two different cognitive processes when transforming representations: transforming representations in one register (such as creating a graph based on a table, called *treatments*) and between registers (i.e., creating a function graph based on an equation, called *conversions*). This review considers only the first type of transformation.

### Graphs of Numerical Data

Graphical representations of numerical data (graphs) are generally based on conventions, such as depicting the dependent variable on the y-axis (Lachmayer et al., 2007), which is particularly important for their function of conveying information (de Vries & Masclet, 2013). Graphs are widespread and frequently used representations that display the variance of a single variable (univariate) or the dependence of two (bivariate) or more (multivariate) variables (Eichler & Vogel, 2012). Numerical data often represent an observed phenomenon, including randomly caused variances, and functional graphs are used to model the structure of a bivariate (or multivariate) dataset via certain mathematical functions (Eichler & Vogel, 2012; Engel, 2018). Interpreting unknown graphs is an important skill because it denotes an act of learning – moving between abstract and concrete representations (Roth & Hwang, 2006). Functional graphs, which in a narrow sense are considered pure mathematical objects without contextual information, are not part of this review because such graphs can be constructed without numerical data using a mathematical equation alone. Graphing draws information from another source of numerical data (e.g., creating a line graph based on a data table). Generating such graphical representations is assumed to improve understanding, as the generation process requires engaging deeply with the structure of the represented numerical data, and the relevant characteristics of datasets can be communicated very effectively via graphical representations (Ainsworth et al., 2011). Therefore, we include studies on *self-generated* or *partially completed* graphical representations and exclude those that exclusively address the interpretation of or learning with *provided* graphs in our review.

Although numerous scientific studies have examined student graphing, to the best of our knowledge, scholars have not conducted a systematic review to synthesize the findings. The present review aims to synthesize the existing results, both positive and negative, regarding the construction of convention-based graphical representations based on numerical data.

## Graphing in STEM Education

Graphing competence covers both the interpretation and creation of graphical representations (Glazer, 2011). Graphing competence is necessary for STEM learning—for example, students should be able to relate the data depicted in a graph to the phenomenon that it describes (Glazer, 2011). The definition of graphing competence also incorporates such concepts as "graph comprehension" (e.g., Curcio, 1987; Kanzaki & Miwa, 2012; Zacks & Tversky, 1999), "graph interpretation" (e.g., Biehler, 2006; Boels et al., 2019; Ergül, 2018; Gaona et al., 2021; Lachmayer et al., 2007; Nixon et al., 2016; Roth & Bowen, 2001), and selecting an appropriate graph type for a certain task (e.g., Baker et al., 2001; Kanzaki & Miwa, 2012; von Kotzebue et al., 2014). The term "graphical literacy" is also related to graphing competence and includes the skill of constructing graphs (Subali et al., 2017); in addition, it influences graph comprehension (Freedman & Shah, 2002).

### Theoretical Framework of Graphing

Several theoretical approaches can be employed to predict the positive effects of graphing on learning. These approaches can be distinguished in terms of two aspects of graphing: graphing as learning using multiple representations (see above) and graphing as an active generative activity. The first perspective assumes that learning with more than one representation facilitates learning, as is the case when creating a second (graphical) representation that complements the first (often numerical) representation (Mayer, 2005, 2014; Schnotz, 2005). Theories on active generative activities argue that learners are more engaged when they generate representations (Chi & Wylie, 2014).

An educationally relevant aspect of graphing is that, for learners, it is an active generative process. Consequently, being located at the high end of Chi and Wylie's (2014) *Interactive, Constructive, Active*, and *Passive* (ICAP) framework, graphing has several advantages. The ICAP framework states that the effectiveness of learning activities decreases from interactive to constructive and from active to passive activities as learner engagement declines. According to the ICAP framework, generating a graphical representation that learners have not seen during learning and that is solely based on numerical data is a constructive activity and should be associated with considerable benefits.

This is supported by generative learning theory, in which Wittrock (1974) suggested that understanding is closely related to generation. He assumed that there are "organizational structures for storing and retrieving information" (p. 182) as well as mechanisms for retrieving stored (prior) knowledge and integrating new material. Wittrock (1992) explicitly stated that the aim of learning is not to store information but to build meaningful connections. This includes paying attention to and making sense of the learning material (Wittrock, 1992). Accordingly, instruction should focus on teaching learners to generate meaning. The overall benefit of generation activities, such as finding synonyms instead of merely reading a text, has also been called the "generation effect" and can be found in literature reviews (e.g., Bertsch et al., 2007). In their meta-analysis, Bertsch et al. (2007) found evidence that the generation effect increased with higher mental effort. When performing an active generation activity during learning, such as constructing a graphical representation based on the provided information, close attention should be paid to the presented information—for example, to the structure of the provided numerical data. This information is then connected to prior knowledge, such as previously learned conventions for interpreting graphical representations. Based on this research, one can assume that graphing is more advantageous to learners than merely looking at graphs and practicing graph comprehension. However, it is difficult to determine the effectiveness of graphing as an instructional method in general based on individual studies, especially because the way graphing is conducted (e.g., manually or via tools) might also influence the learning outcome.

There are numerous ways in which graphing activities can be implemented in education. In our review, we are interested in not only how graphing has been examined in studies but also whether and how authors embedded graphing into existing theories. As theoretical contextualization may depend on the purpose of education, analyzing the possible implications of this link could produce valuable insights for educators. The purpose and advantages of graphing in the STEM context may differ from the results found regarding the generation of external representations in general.

### *Previous Research on Graphing*

Previous literature reviews have focused on students "drawing" or "sketching" during learning (Cromley, Du, & Dane, 2020; Fiorella & Zhang, 2018; van Meter & Garner, 2005; Wu & Rau, 2019), which, as generative processes, have certain similarities to graphing. According to van Meter and Garner (2005), sketches of various activities, such as swimming (Schmidgall et al., 2019), are representations of a mental image; therefore, the designs of representations can vary between learners. However, graphing follows general construction conventions, which means that graphing, when done correctly based on numerical data for a specific purpose, leads to comparable results regardless of the learner. Despite this, graphing is not a simple procedure, as even scientists sometimes do not choose the most appropriate graph type to display their data and, for example, seem to prefer line graphs to other types of graphs (Weissgerber et al., 2015).

The generation of representations is used to visualize and externalize information (Schmidgall et al., 2019). According to Stern et al. (2003), active graph generation supports an understanding of how graph design is generally related to the contents represented, compared with just exploring a given graph. This general understanding is transferable across domains. Furthermore, Stern et al. (2003) found that due to an enhanced examination of the material (generation effect), learners benefitted from generating graphs during a problem-solving task involving graphing in stock-keeping. In her review of graph interpretation, Glazer (2011) recommended instructing students in graph interpretation and graph creation (see also Cox, 1999). Such focused instruction can guide the generation of graphs, improve their accuracy, and highlight key points in graph design as well as the relationships between graphical representations (van Meter & Garner, 2005). Support during graphing can facilitate comprehension, although generation increases students' cognitive load compared to learning with provided representations (Zhang & Fiorella, 2021). Generating external representations also improves problem-solving (Cox, 1999), deep understanding, and knowledge transfer (Chi & Wylie, 2014).

In educational practice, generating graphical representations is relevant to several disciplines, most notably STEM subjects. Díaz-Levicoy et al. (2018) evaluated statistical graphs in mathematics textbooks for first- to sixth-grade students in primary education and found that generating graphs was the second most frequently taught activity (after basic mathematical operations, such as addition). A previous review by Leinhardt et al. (1990) of the literature on graphing and graph interpretation of functions also highlighted the importance of the ability to construct graphs because it "can be seen as one of the critical moments in early mathematics" (p. 2). Convention-based graphical representations of numerical data are used not only in mathematics but also in other contexts, such as physics, and research on graphing focuses on diverse learning objectives.

Using a pretest-posttest design, Mevarech and Kramarsky (1997) noticed a positive effect of graphing instruction on eighth-grade students' manual graphing abilities but also identified persistent difficulties. In a three-year-long study that analyzed students' reasoning when constructing line graphs, Wavering (1989) found that students' reasoning ability increased over time. This is particularly important because graphing is a relevant aspect of scientific inquiry (Gooding, 2010). For example, Schultheis et al. (2023) used authentic research experiences (Data Nuggets), which included graphing authentic datasets, in the context of biology and found that students improved in using scientific constructing scientific explanations. They were also more interested in STEM careers. Bahtaji (2020) compared undergraduate students' learning in the following three conditions, examining the conditions' effects on conceptual knowledge and graphing skills: (a) with provided graphs, (b) with self-constructed graphs, and (c) with self-constructed graphs made using explicit graphing instruction. He found that although all interventions facilitated conceptual knowledge acquisition, only explicit instruction developed graphing skills. Similarly, Harsh and Schmitt-Harsh (2016) used an instructional design to enhance graphing skills during a general education science course at a university and found that students improved between the pretest and the posttest. Angra and Gardner (2016) discovered differences between novices and experts in the construction of graphical representations based on tables. Along the same lines, experts' explanations were consistent with the graphs that they generated, which was not necessarily the case for novices (Kanzaki & Miwa, 2012). Based on a qualitative analysis, Angra and Gardner (2016) developed a step-by-step guide for teaching students how to choose a graph type, construct the graph after planning each step, and, finally, critically reflect on the graph choice. Based on these steps, teachers can adapt their instructions in relation to students' answers, and students can develop a clear process for graphing.

Furthermore, graphing can be performed manually as well as with the help of software tools. For example, in a qualitative analysis, Parnafes and Digoodi (2004) found that reasoning type was related to representation type when middle school students interacted in groups with the software environment NumberSpeed. The students used content-based reasoning when working with number–list representations and more model-based reasoning when working with spatial and dynamic motion representations. Nixon et al. (2016) noted that undergraduate students' understanding of best-fit lines changed depending on the physics lab activity. Biehler's (2006) results indicated that students had difficulties interpreting graphical representations in the context of a task during a qualitative analysis of students' group work with Fathom software.

Overall, previous studies in STEM education have revealed (a) a variety of educational settings in which the graphing of numerical data is useful (Bahtaji, 2020; Harsh & Schmitt-Harsh, 2016), (b) the cognitive aspects necessary for and during graphing (e.g., Mevarech & Kramarsky, 1997; Wavering, 1989), and (c) students' graphing strategies and their interpretations in the context of physics lab activities (Nixon et al., 2016). In the present review, we aim to systematically synthesize and summarize studies on the graphing of numerical data to facilitate the comparison of various methods across different STEM contexts in terms of their similarities, differences, and effectiveness.

### *Students' Difficulties*

Understanding students' difficulties is important to facilitate graphing competence. In a literature review, Boels (2019) described conceptual difficulties when interpreting histograms related to the interpretation of data and distribution. Moreover, von Kotzebue et al. (2014) investigated the mistakes made by 437 science students when constructing diagrams in the context of biology. Students had trouble in all analyzed categories, such as choosing the correct diagram type, assigning variables to axes, and scaling. Nixon et al.

(2016) observed that determining the axes' scaling posed a problem for students when graphing during physics lab activities. Baker et al. (2001) asked eighth and ninth graders to select, interpret, and construct graphs. Most students' performance seemed to depend on their ability to transfer knowledge from bar graphs to other graphical representations of data. Bayri and Kurnaz (2015) examined eighth-grade students who constructed different types of graphical representations and concluded that students had difficulty shifting between different types of representations. The authors assumed this to indicate that, for students, various types of graphical representations could have different meanings. One reason for this could be that students see representations as analogue reflections of reality rather than representations of a symbolic character (graph-as-picture error; Clement, 1989).

These studies show how difficult graphing can be for students and how important it is for instructors to be aware of these difficulties. Therefore, it is important to consider not only how graphing is implemented in various studies and its effectiveness in these contexts but also the possible problems faced by students.

## Research Questions

Based on previous research, our literature review aims to systematically examine graphing in STEM subjects in K-12 education. This is the context in which students learn how to construct graphs and where graphing is often assessed (i.e., in exams). This review therefore addresses the following questions:

(1) How is graphing implemented in studies on this activity in K-12 STEM education?

(2) What is the effectiveness of graphing as an instructional method in K-12 STEM education?

(3) Which difficulties can arise when graphing in K-12 STEM education?

# Methods

This systematic review is documented and reported according to the guidelines for preferred reporting items for systematic reviews and meta-analyses 2020 (PRISMA, Page et al., 2021). In the following section, we present the exact method.

## Inclusion and Exclusion Criteria

An article's suitability was judged based on the inclusion criteria. We were interested in all studies that examined the graphing in K–12 STEM education. Participants were students in K-12 education who participated in studies with a STEM topic. We were only interested in self-generated or partially completed graphical representations rather than pure graph interpretations. The representations should be based on convention; completely new or made-up representations were not included. Furthermore, the generated graphs should be based on numerical data. Studies that investigated graphing based on given verbal–textual or mathematical–symbolical information were excluded. All criteria are summarized in Table 1.

**Table 1**

*Criteria for Inclusion and Examples of Exclusion*

| Criterion | Inclusion | Exclusion examples |
|---|---|---|
| Subjects | K-12 students | University students |
| STEM Education | STEM topic | Liberal arts topic |
| Graphing | Self-produced graph | Provided graphs |
| STEM practice | Convention-based graph | Self-invented representations |
| Numerical data | Data-based graph | Function graph not based on data |
| Publication language | English | Korean |
| Publication type | Empirical research | Theoretical research |
| Scientific evaluation | Peer review | Grey literature |

**Literature search**

The search was conducted in March 2022 using the following three major scientific databases selected by our interdisciplinary review team due to the high quantity of educational research in various disciplines: Scopus, ERIC, and PsycInfo. ERIC and PsycInfo were accessed via EbscoHost. Relevant terms in the categories of Education, Graphing, and Data use were based on previous scoping searches to form a complete yet economical search term. The full search term was *(educat\* OR student\*) AND (graphing OR graph OR graphs OR plotting OR plot OR plots) AND (data\* OR variable\* OR construct\*)*. We considered the title, abstract, and keywords in the systematic search.

Exported records were saved, managed, and deduplicated using the reference management software Mendeley. After deduplication, the records were screened. In the first step of the screening, only the title and abstract were considered. For this, we used ASReview (Utrecht University, 2021). ASReview speeds screening by using machine learning to prioritize the articles found during the systematic search (van de Schoot et al., 2021). The software continuously updates the order of the titles and shows the most relevant studies first. The algorithm was first trained using included and excluded articles, which were found during scoping searches and discussed by the review team. Each user decision (include/exclude) made in ASReview was used to improve the model during the entire screening process. Two coders from the review group trained the algorithm up to a termination criterion (<5 relevant titles in 100). The same two raters then manually screened the studies that were judged relevant when considering the full text using the software

Rayyan (Ouzzani et al., 2016). Any discrepancies were solved via discussions between the two raters and, if necessary, with the broader review group. Therefore, we reached complete agreement between the two raters. In the next step, relevant conferences in the psychological and educational contexts were reviewed for the relevant proceedings according to the following eligibility criteria: We looked for international conferences in a (STEM) education context, with English abstracts, and full papers published as part of the proceedings. The conferences should also be peer-reviewed. Seven conferences matched these criteria: the International Conference of the Learning Sciences (ICLS), the International Conference on Teaching Statistics (ICOTS), and the meetings of the American Educational Research Association (AERA), the European Association for Research on Learning and Instruction (EARLI), the European Science Education Research Association (ESERA), the Cognitive Science Society, and the National Association for Research in Science Teaching (NARST). We restricted our search to start from the year 2019 because we assumed that relevant research would be published as a paper that would be found in the normal search process. For articles included after the full-text screening, a forward and backward search for further potentially relevant articles was conducted. A second search for papers published after the initial search was conducted in April 2024 to update the results. A complete overview of the literature search and screening process can be seen in Figure 1.

**Figure 1**

*PRISMA Flow Diagram Detailing the Study Selection*

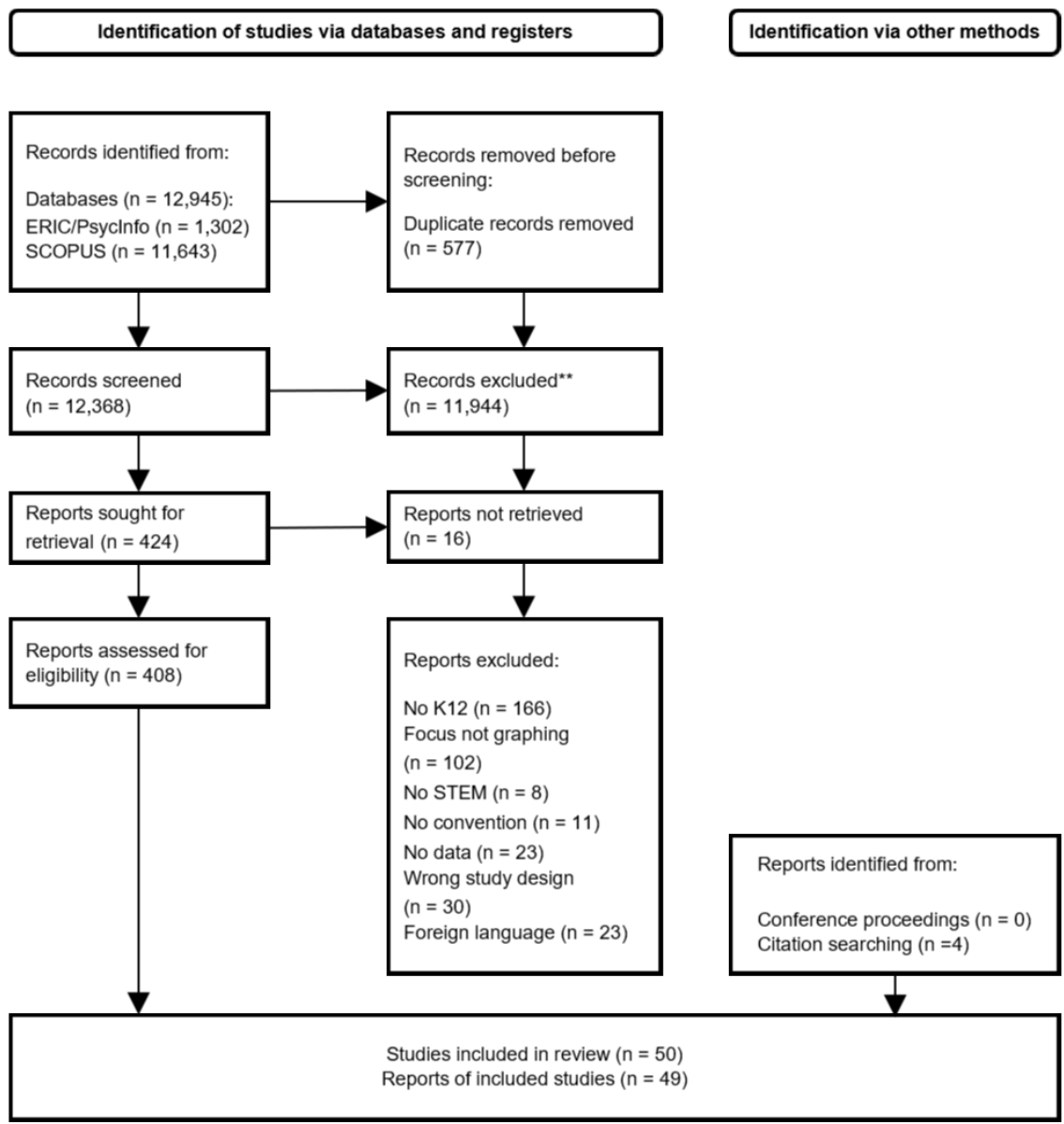

**Data Extraction and Coding Schemes**

In our literature review, we summarize and synthesize studies that deal with the contexts, effects, and difficulties of graphing in K–12 STEM education. For each included study, we extracted relevant data (see Table 2 for an overview of example codes). Two coders coded all studies, and possible disagreements were resolved via discussion.

We extracted information about the sample and the general study characteristics to structure the body of the research. We classified the selected studies according to the addressed topic, as research indicates that students' representation use and graph interpretation can differ between domains and contexts (Chang, 2018; Roth & Bowen, 2001). Furthermore, we are interested in the types of graphical representations that students generated to compare the graphing of univariate (e.g., bar graphs) and multivariate data (e.g., scatter plots). This could be relevant because, for example, bar and line graphs are associated with different concepts, such as discrete comparisons and trends (Zacks & Tversky, 1999). For graphing, it may be important whether the learners generated the data themselves, as described by Nixon et al. (2016), or whether they were provided with the data (e.g., Bahtaji, 2020). This might be relevant as self-generated and provided data have different learning benefits (Hug & McNeill, 2008). With the availability of computers and other technical tools, there might be differences due to technological difficulties (Chang et al., 2024) between manual graphing (e.g., Wavering, 1989; Subali et al., 2017) and the construction of graphical representations via programs, whose availability has increased in recent years (e.g., Kohnle et al., 2020; Lee & Lee, 2018). This might influence the perspective from which students view data because it might be easier to create representations of aggregate data, such as box plots, with tools than manually (Konold et al., 2015). Moreover, we would like to know whether a gender-balanced study design is comparable to, for example, an unbalanced design (e.g., Castro-Alonso et al., 2019). For each included study, the ratio of male to female participants was coded. This allowed us to assess whether gender was considered in the research, the extent to which the effects found might be generalized across genders, and whether participants' gender was a possible moderator of the effects. All themes, categories, and example codes for the coded variables are shown in Table 2. If variables were not reported, they were coded as "NR."

**Table 2**

*Themes, Categories, and Example Codes From the Data Analysis*

| Theme | Category | Example Code |
|---|---|---|
| Population | Children | Pre-school, primary school (approx. age 5–10), high school (age >10/11) |
| General information | Gender | Balanced, unbalanced (female skew), unbalanced (male skew) |
| | Country | Germany, the USA, the Netherlands |
| | *N* | 92,110 |
| Topic | STEM | Biology, chemistry, physics, computer science, engineering, math, technology |
| Graphing | Method | Manual, tool-based |
| | Graph type | Histogram, box plot, line graph, scatter plot |
| | Guidance | Minimal, explicit, completing, comparison |
| Numerical data | Collection | Self-generated, given |
| | Data type | Univariate, bivariate, multivariate |
| Study design | Activity | Problem-solving, experiment, instruction |
| | Study setting | Lab, lesson, course |
| | Moderating variable | Prior knowledge, motivation, spatial, load |
| Results | Analysis | Quantitative, qualitative, mixed methods |
| | Overall effect of graphing | Positive, negative, inconclusive |
| | Effect size | *r, d* |
| | Hypothesis | Confirmed, unclear |

| Theory | Cognitive theory for graphing | CTML/ITPC, generative learning, ICAP, other[a] |
|---|---|---|
| Difficulties | Construction difficulties | Graph construction; Variable ordering; Data translation |
| | Theoretical difficulties | Connection to concepts; Interpretation |

*Note.* CTML: Cognitive theory of multimedia learning (e.g., Mayer, 2014), ITPC: Integrated model of text and picture comprehension (Schnotz, 2005), ICAP: Interactive, active, constructive passive framework (Chi & Wylie, 2014).

### Data Analysis

The data extracted from the included studies were saved and processed in Microsoft Excel and Access. In addition to answering the research questions, we performed a narrative comparison of the studies in terms of their use of univariate versus multivariate data.

Publication bias is indicated by the systematic difference between published and unpublished research (Vevea et al., 2019). As this is, to the best of our knowledge, the first literature review of graphing in STEM subjects, we wanted to ensure high quality by only including peer-reviewed studies. We believe that bias was sufficiently reduced by using multiple search engines and covering various topics and journals, as well as by including proceedings from conferences judged relevant by experts in various STEM fields.

### Results

In the following sections, we present the results according to the codes (see Table 2). Sometimes, the authors did not explicitly mention the information encoded for our review; these studies are not reported below. An overview of all the studies included in the literature review can be seen in Table 3. A complete overview of all codes is provided in the tables of the supplementary material.

**Table 3**

*Overview of the Relevant Codes for the Included Studies*

| Authors | Year | STEM Topic | Data Collection | Construction Difficulties | Theoretical Difficulties |
|---|---|---|---|---|---|
| Åberg-Bengtsson | 2006 | Math | Both[c] | GC[e] | Part vs. Whole |
| Adams & Shrum | 1990 | Biology | SG[d] | NR | NR |
| Arteaga et al. | 2020 | Math | Both | GC | NR |
| Ates & Stevens | 2003 | Chemistry | SG | NR | NR |
| Aydın-Güç et al. | 2022 | Math | Given | DT | IP[h] |
| Berg & Phillips | 1994 | NR[a] | Given | GC | Concept |
| Branisa & Jenisova | 2015 | Chemistry | Given | GC, DT[f] | IP, Concept |
| Brasell & Rowe | 1993 | Physics | Given | GC, VO[g] | Concept |
| Detiana et al. | 2020 | Math | SG | NR | NR |
| Dewi et al. | 2018 | Physics | NR | GC, VO | IP, Concept |
| Dimas et al. | 2018 | Physics | Given | GC | Concept |
| English & Watson | 2015 | Math | SG | GC | NR |
| English | 2022 | Physics | SG | NR | Concept |
| English | 2023 | Math | SG | NR | Concept, Part vs. Whole |
| Fielding-Wells | 2018 | Math | SG | NR | IP |
| García-García & Dolores-Flores | 2019 | Math | SG | DT | NR |
| Garcia-Mila et al. (Study 1) | 2014 | Math | Given | GC, VO, DT | NR |
| Garcia-Mila et al. (Study 2) | 2014 | Math | Given | GC, VO | NR |
| Gardenia et al. | 2021 | Math | SG | DT | NR |
| Gerard et al. | 2012 | NR | SG | NR | Concept |
| Gultepe & Kilic | 2015 | Chemistry | Given | GC, VO | NR |
| Gültepe | 2016 | Chemistry | NR | DT | IP, Concept, GT[i] |
| Harrison et al. | 2019 | NR | SG | VO | IP |
| Jackson et al. | 1992 | CS[b] | Given | GC | GT |
| Jackson et al. | 1993 | CS | Given | GC | IP, GT |
| Karplus | 1979 | Math | Given | VO | Concept |
| Kramarski | 1999 | Math | SG | GC, DT | NR |
| Meisadewi et al. | 2017 | Biology | NR | NR | NR |
| Mevarech & Kramarsky | 1997 | Math | SG | GC, VO | NR |
| Moritz | 2003 | Math | Both | GC, VO, DT | NR |
| Ng & Nicholas (Study 2) | 2011 | Biology | SG | NR | NR |
| Nurrahmawati et al. | 2021 | Math | SG | VR, DT | IP |
| Onwu | 1993 | Various | Given | GC, VO | IP |
| Oslington et al. | 2020 | Math | Both | GC, DT | Concept |
| Ozmen et al. | 2020 | Math | Given | GC, VO | GT |
| Padilla et al. | 1986 | Various | Given | GC, VO | NR |
| Pols | 2019 | Physics | SG | GC | Concept |
| Pospiech et al. (Study 1) | 2019 | Physics | Both | NR | NR |
| Pratt | 1995 | Math | SG | VO | IP |
| Rahmawati et al. | 2020 | Math | SG | NR | NR |
| Saldanha & McAllister | 2016 | Math | Given | NR | NR |

| Stephens | 2024 | Physics | SG | NR | IP, Concept |
| Struck & Yerrick | 2010 | Physics | SG | VO | NR |
| Tairab & Al-Naqbi | 2004 | Biology | Given | DT | IP |
| Vitale et al. | 2019 | Physics | Both | NR | Concept |
| Watson | 2022 | Physics | SG | NR | IP |
| Watson et al. | 2023 | Physics | SG | GC | NR |
| Wavering | 1989 | Math | Given | GC | NR |
| Webb & Boltt | 1991 | Biology | Given | NR | IP |
| Wu & Krajcik | 2006 | Various | SG | GC | IP |

*Note:* [a] NR: None reported, [b] CS: computer science, [c] given as well as self-generated data: Both, [d] self-generated: SG, [e] graph construction: GC, [f] data translation: DT, [g] variable ordering: VO, [h] interpretation: IP, [i] graph type: GT

### General Information

The studies included in the review were published between 1979 and 2024, which is the last year in which we identified relevant articles in the literature review (see Figure 2). The rate of publications on graphing seems to have reached its maximum in the years 2019 and 2020.

**Figure 2**

*Overview of the Publication Years of the Included Studies*

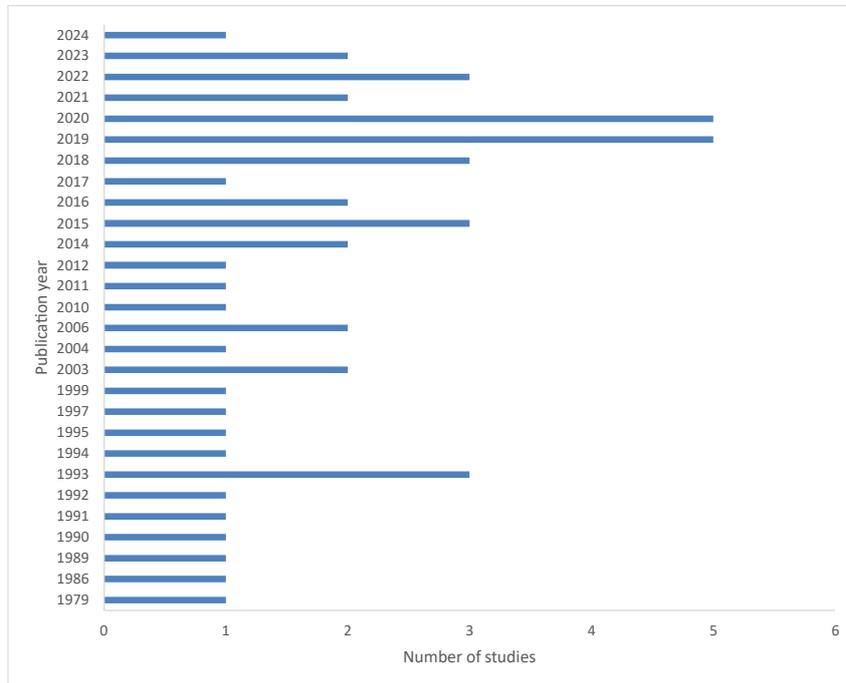

Most of the studies were conducted in the USA, followed by Indonesia and Australia (see Figure 3). Few studies were published in Europe, which is surprising because other studies reference, e.g., the German curriculum (Meisadewi et al., 2017).

**Figure 3**

*Overview of the Countries Where the Studies were Published*

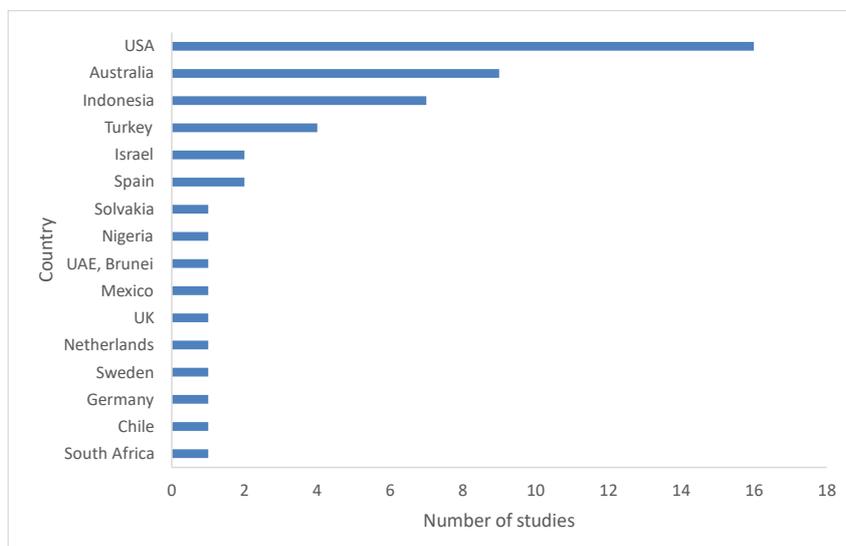

Most studies did not base their research on a theory specific to graphing (n=23). Two studies referred to generative learning (Åberg-Bengtsson, 2006; Vitale et al., 2019). Additionally, 27 studies referred to other theories. For example, Ainsworth's (2006) design, function, and task (DeFT) framework, which is based on the functions that representations have during instruction, was referenced for this purpose (Dimas et al., 2018), but it was also used in one study as a theoretical basis for translating between two types of representations, such as the "mathematical representation translation from verbal to graph" (p. 401, Rahmawati et al., 2020). Similarly, English (2022) referred to metarepresentational competence and the need to connect graphical representations to the represented concept (see also Vitale et al., 2019), as well as transform types of representations (Garcia-Mila et al., 2014). Other tool-based studies have argued that computer-based graphing frees up cognitive resources (Jackson et al., 1992, 1993). Another perspective includes a theoretical background based on the progression of logical thinking abilities indicated by graphing (Berg & Phillips, 1994). Graphing has also been used as a good research tool to capture "students' thinking processes" (Ng & Nicholas, 2011, p. 79).

Another study used active learning as its theoretical basis (Pratt, 1995). One study used a pedagogical approach called *Active Graphing* (Åberg-Bengtsson, 2006), introduced in previous research (Ainley, 2000). For microcomputer-based laboratories (MBLs), authors based their research on the benefit of having a "genuine scientific experience" (p.778, Adams & Shrum, 1990) or on the view of "graphing as practice" (p.57, Ates & Stevens, 2003) and did not consider graphing from a cognitive perspective. This scientific (Fielding-Wells, 2018; Stephens, 2024; Watson et al., 2023; Wu & Krajcik, 2005) and practical (Aydın-Güç et al., 2022; Gültepe, 2016; Harrison et al., 2019; Oslington et al., 2020) view is stated similarly in other studies, for example, as part of problem-solving (Bransia & Janisova, 2015; Pospiech et al., 2018).

## Implementation of Graphing

### *Study Design and Context*

Most authors analyzed graphing in a problem-solving environment (n=36), although participants received graphing instruction during some studies (n=6). For example, Mevarech and Kramarsky (1997) asked 92 eighth-grade students to construct graphs before and after a graphing unit and qualitatively analyzed their progress based on the students' responses. Several studies investigated graphing in the context of experimentation (n=10), such as investigating tool-based graphing in the context of oscillation using a spring-mass simulation (Stephens, 2024). Meisadewi et al. (2017) found that lab-based activities could improve students' graphing skills [see also Struck & Yerrick (2009) for similar results, as well as Gerard et al. (2012) for a mixed-methods analysis]. Three studies used an instructional context combined with problem-solving (Åberg-Bengtsson, 2006; Gerard et al., 2012; Vitale et al., 2019), for example, Åberg-Bengtsson (2006) instructed elementary students on how to use the software Excel in a collaborative setting and investigated their reasoning during graph construction.

Studies were most often conducted in more than one lesson (n=22), followed by a single lesson (n=17) and interviews (n=5). Interviews were sometimes conducted concurrently with lessons (Aydın-Güç et al., 2022; Karplus, 1979); for example, Karplus (1979) asked 414 high-school students to solve "functionality puzzles" (p. 398) during a lesson and investigated the answers of 37 students during interviews to "clarify the written answers" (p. 398). Only one study was conducted in the researchers' lab (Pospiech et al., 2019).

The number of participants differed greatly between studies, from three (Detiana et al., 2020; Gardenia et al., 2021) to 745 (Arteaga et al., 2020). On average, 116 students participated (SD = 174). Students in high school participated in most studies (n=43), with only six studies analyzing graphing in primary schools. One study analyzed students from grades three to seven, spanning both primary and high school (Moritz, 2003). The gender of the students was often not reported (n=31). Six studies had an equal number of female and male participants and in nine studies more males than females participated. There were four studies where more females than males participated.

The STEM topics varied between studies (see Figure 4). Graphing was mostly conducted on the topic of mathematics (n=22); for example, Moritz (2003) examined how primary and high-school students constructed coordinate graphs during their mathematics lessons. Other studies were conducted in biology, physics, chemistry, and computer science. In three studies, students constructed various contexts (Onwu, 1993; Padilla et al., 1986; Wu & Krajcik, 2006) — for example, by "incorporat[ing] fundamental science concepts across several science disciplines" (p.66, Wu & Krajcik, 2006).

**Figure 4**

*STEM Topics Used for Graphing*

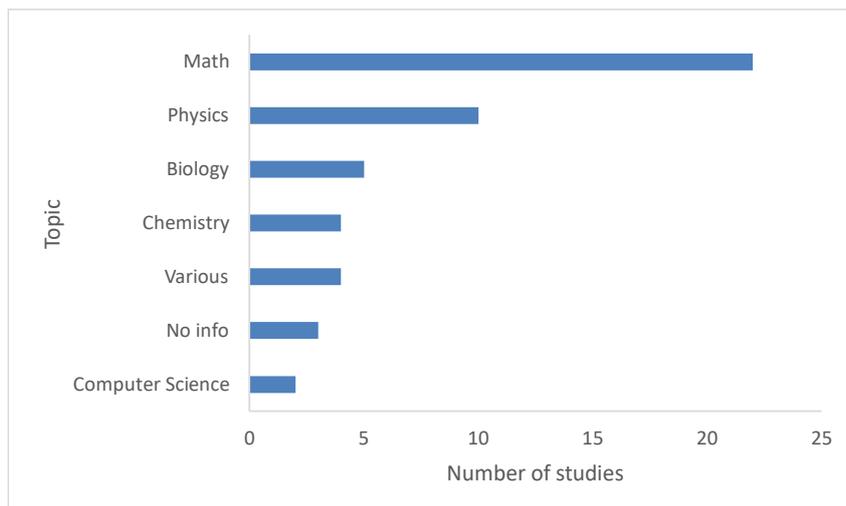

*Graph Types, Graphing Method, and Graphing Guidance*

Students constructed numerous types of graphs (see Figure 5). The most frequently constructed graph type was a line graph (n=27). Other studies let students decide the type of graphs, e.g., for a task during a test (Ozmen et al., 2020). Bar graphs and tables were also common types of graphs. Two studies did not specifically state which types of graphs they used (Branisa & Jenisova, 2015; Meisadewi et al., 2017).

**Figure 5**

*Types of Graphs Constructed by Participants*

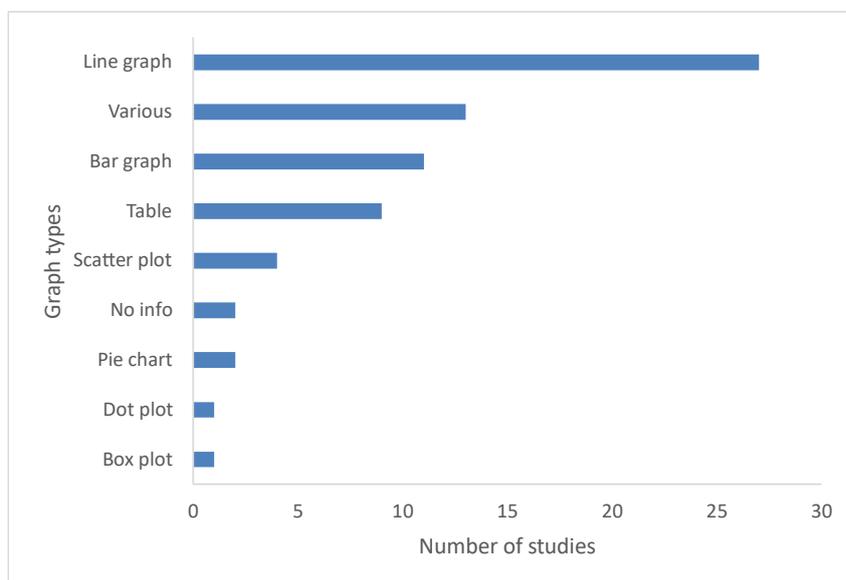

*Note:* Multiple mentions are possible.

During most studies, participants created graphs by hand (n=29); for example, Nurrahmawati et al. (2021) investigated high-school students' errors when translating between mathematical representations, such as between tables and line graphs. Nine studies analyzed graphs created via tools (Gerard et al., 2012; Harrison et al., 2019; Jackson et al., 1992; Jackson et al., 1993; Ng & Nicholas, 2011; Pratt, 1995; Saldanha & McAllister, 2016; Stephens, 2024; Vitale et al., 2019). Ten studies examined graphs created both manually and via tools (Åberg-Bengtsson, 2006; Adams & Shrum, 1990; Ates & Stevens, 2003; Brasell &

Rowe, 1993; English & Watson, 2015; English, 2022; Kramarski, 1999; Watson et al., 2022; Wu & Krajcik, 2006). All studies examining tool-based graphing used various computer applications, such as TinkerPlots (English & Watson, 2015; Saldanha & McAllister, 2016; Watson et al., 2022; Watson et al., 2023), the web-based inquiry science environments WISE (Gerard et al., 2012; Harrison et al., 2019; Vitale et al., 2019), the web-based data exploration environment CODAP (Stephens, 2024), CricketGraph (Jackson et al., 1992, 1993), or Excel (Åberg-Bengtsson, 2006; English, 2022).

Participants often did not receive any guidance about how to create graphs, although ten studies provided explicit instructions (see Figure 6). For example, Detiana & Mailizar (2020) showed high-school students instructional video tutorials on function graphs before asking them to manually construct their own graphs. Participants received minimal instruction in five studies (Brasell & Rowe, 1993; Jackson et al., 1993; Karplus, 1979; Pratt, 1995; Webb & Boltt, 1991), which included providing participants with worksheets, including frameworks for the graphs (Webb & Boltt, 1991). In one study, the students were asked to complete a representation consisting of a table with corresponding graphs (Åberg-Bengtsson, 2006). In two other studies, students were asked to compare their representations to graphs created by fictional students (Harrison et al., 2019) or to compare their graphs with teacher-generated graphs constructed with TinkerPlots (Watson et al., 2022). Guidance was coded as "student-based" for a study in which participants were interviewed while they constructed graphs and received feedback based on their progress (García-García & Dolores-Flores, 2019).

**Figure 6**

*Types of Guidance for Graphing*

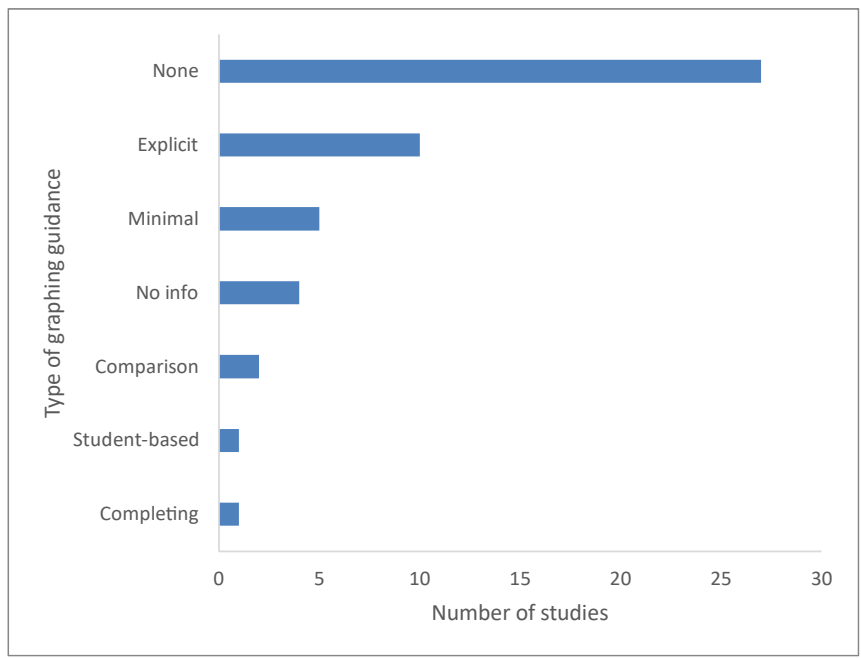

## Numerical Data Used for Graphing

The data that the students used for graphing varied. Participants mostly created the data themselves (n=23), either during experiments by measuring variables or by creating the data for graphing, e.g., based on problem statements (Gardenia et al., 2021). Eighteen studies provided participants with the data to create graphs, and some studies used a combination of both (n=6). For example, Oslington et al. (2020) provided third-graders with temperature data and asked them to predict future temperatures before constructing a graphical representation of the data.

Most of the numerical data used for graphing was bivariate (n=34), compared to univariate (n=6) and multivariate (n=3) data. Two studies provided students with a variety of data; for example, Jackson et al. (1992) provided students with graphing instruction for *Cricket Graph*^TM in a computer science course using a variety of contexts and data.

**Effectiveness of Graphing as an Instructional Method**

*Study Results and Moderators*

Most studies reported no hypothesis (n=45) and no effect sizes (n=49) or results regarding the benefits of graphing or their instruction (n=38). This might be due to the number of qualitative (n=22) and mixed-methods analyses (n=19). Nine studies analyzed their data quantitatively (Adams & Shrum, 1990; Ates & Stevens, 2003; Harrison et al., 2019; Meisadewi et al., 2017; Onwu, 1993; Padilla et al., 1986; Struck & Yerrick, 2010; Wavering, 1989; Webb & Boltt, 1991).

Eleven studies reported positive effects (Adams & Shrum, 1990; Branisa & Jenisova, 2015; Karplus, 1979; Gerard et al., 2012; Gultepe & Kilic, 2015; Meisadewi et al., 2017; Mevarech & Kramarsky, 1997; Padilla et al., 1986; Struck & Yerrick, 2010; Vitale et al., 2019; Wu & Krajcik, 2006). Two studies reported positive effects for manual graphing (Adams & Shrum, 1990; Branisa & Jenisova, 2015). Branisa and Jenisova (2015) compared manual graphing with automatically generated graphs and found that students practiced manual graph construction based on experimental data performed better than students who were provided with computer-constructed graphs. Adams and Shrum (1990) found that conventional graphing instruction was better than instruction with microcomputers. They were the only authors who reported an effect size (−1.01). Some studies reported positive effects of various types of instruction (Gerard et al., 2012; Gultepe & Kilic, 2015; Meisadewi et al., 2017; Mevarech & Kramarsky, 1997; Struck & Yerrick, 2010; Wu & Krajcik, 2006) on graphing skills. For example, Gerard et al. (2012) compared drawing tools with probe-based tools and found that probe-based tools might be better for learning how not to see graphs as pictures but as representations of data. However, students who drew instead of using motion sensors "constructed more precise graphs and verbal interpretations" (Gerard et al., 2012, p. 569). Another possible moderator of graphing skills might be the level of education because Karplus (1979) and Padilla et al., (1986) reported an increase in skills with a progression between grades. While Karplus (2006) compared students between sixth and eighth grade who constructed function graphs based on data pairs, his results correspond to the results of a mixed-methods approach by Mevarech and Kramarsky (1997), who found that instruction improved students' graphing performance. The goal of instruction varied between studies: some specifically wanted to improve students' graphing skills (Gerard et al., 2012; Meisadewi et al., 2017; Mevarech & Kramarsky, 1997), whereas others taught specific topics, such as kinetics (Struck & Yerrick, 2010) or water quality (Wu & Krajcik, 2006). One study aimed to improve students' scientific argumentation skills (Gultepe & Kilic, 2015). Furthermore, students benefited from graphing data that illustrated their ideas and revising their graphs based on scientific concepts (Vitale et al., 2019). Of the studies reporting positive effects, five analyzed their results quantitatively (Adams & Shrum, 1990; Meisadewi et al., 2017; Padilla et al., 1986; Struck & Yerrick, 2010) and one qualitatively (Wu & Krajcik, 2006). Six studies used a mixed-methods approach (Branisa & Jenisova, 2015; Gerard et al., 2012; Gultepe & Kilic, 2015; Karplus, 1979; Mevarech & Kramarsky, 1997; Vitale et al., 2019). One study reported inconclusive results (Ates & Stevens, 2003). None of the included studies reported any negative effects of graphing.

Most studies (n=37) did not document specific participant characteristics influencing graphing skills. Thirteen studies reported possible moderators. Most effects were reported for age or grade (n=7) and types of mathematical understanding (n=3). For example, Gardenia et al. (2020) found that students with high mathematical skills were better able to construct mathematical representations than students with low or medium skills. Other studies mentioned moderating effects due to cognitive development (n=4), such as reasoning skills (Ates & Stevens, 2014; Berg & Phillips, 1997; Wavering, 1989). One study described results dependent on gender with males performing better than females(Wavering, 1989), maybe due to advanced reasoning skills (Berg & Phillips, 1994).

## Difficulties During Graphing

### *Difficulties with Graphing Conventions*

Most studies reported that their participants had difficulties creating graphs that could be attributed to graphing conventions (n=34). Several studies noted more than one difficulty. As the studies varied extensively, we distinguished between three broad categories of difficulties. The first two types of difficulties are based on structural models that describe the graph construction process (Lachmayer et al., 2007). Graph construction difficulties are concerned with constructing the structure of the graphs, such as scaling the axes and assigning variables to the appropriate axes, whereas variable ordering difficulties describe problems, such as charting points at the correct locations in the graphs. For example, Watson et al. (2023) found that students sometimes had trouble scaling the axes when using TinkerPlots. Onwu (2014) reported that only 38% of the 366 junior high school students had could correctly determine the x- and y-coordinates of data points. An example graph for a variable translation difficulty can be seen in Figure 7. Data translation difficulties were found the least (n=11); they describe difficulties translating from one representation to another. For example, third graders seem to have trouble translating self-constructed tables into suitable graphical representations (Oslington et al., 2020). An overview of students' graphing difficulties due to the conventions can be seen in Figure 8.

**Figure 7**

*Example of a graphical representation with variable ordering difficulties (based on Mevarech & Kramarsky, 1997)*

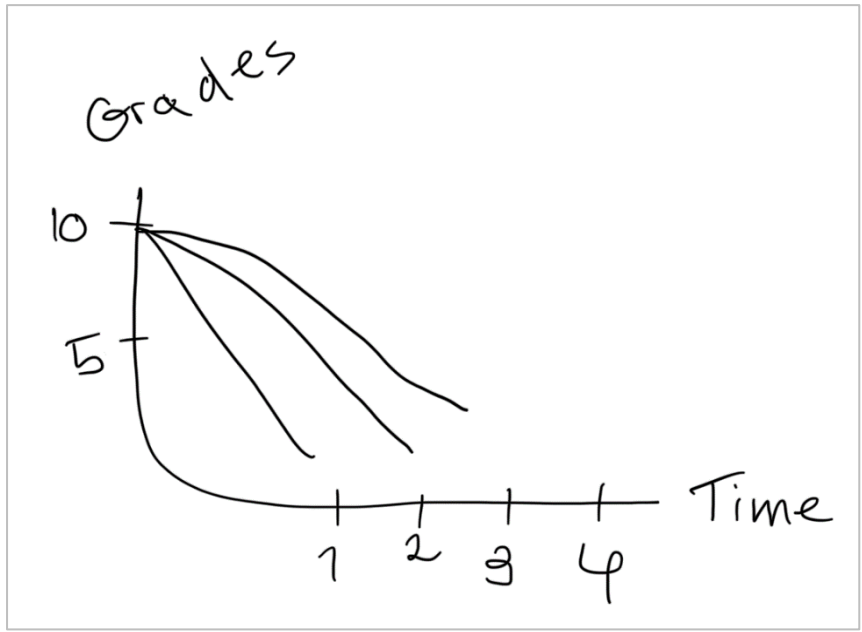

**Figure 8**

*Overview of Students' Difficulties due to Graphing Conventions*

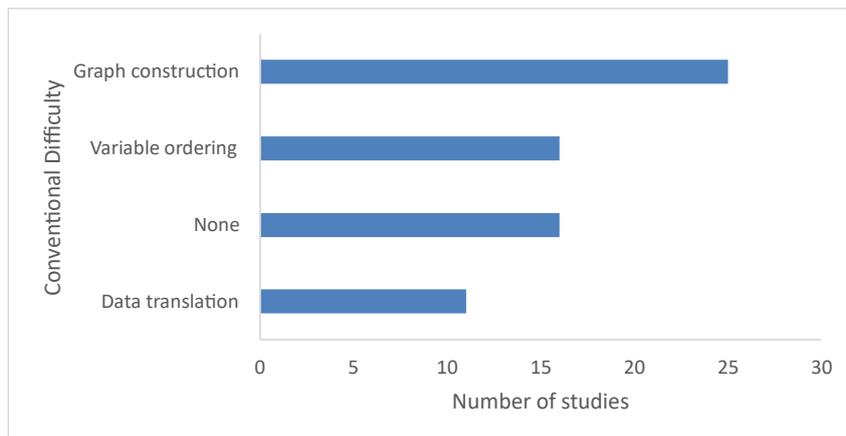

*Note:* Multiple mentions are possible.

### *Theoretical Difficulties During Graphing*

Some students also faced theoretical difficulties (see Figure 9); however, almost half of the studies reported no difficulties (n=28). Conceptual understanding and interpretation were the most common theoretical difficulties described in the included studies. Fourteen studies reported difficulties connecting the graphs to the depicted STEM concept, such as graph-as-picture errors (Gerard et al., 2012). Other students had trouble interpreting the data, for example, to make predictions (Webb & Boltt, 1991). Four studies documented difficulties in choosing the correct graph type for a task (Gültepe, 2016; Jackson et al., 1992; Jackson et al., 1993; Ozmen et al., 2020), although this difficulty seemed to lessen with increasing experience (Karplus, 2006).

**Figure 9**

*Overview of Students' Theoretical Difficulties*

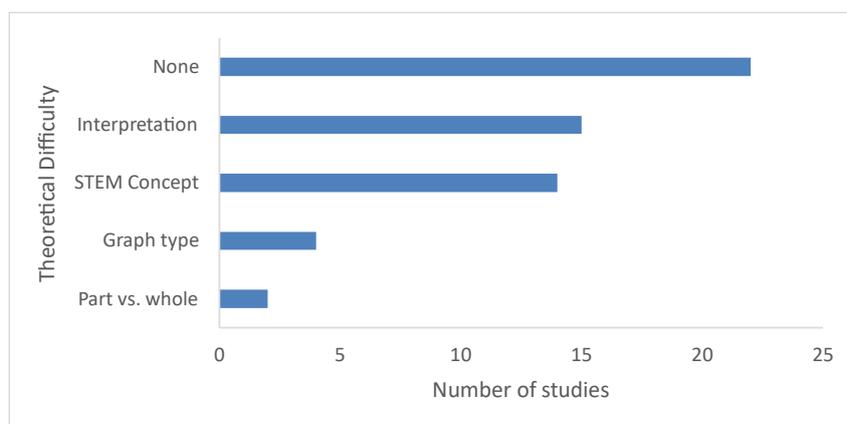

*Note:* Multiple mentions are possible.

Twenty studies reported both theoretical difficulties and difficulties associated with graph construction; for example, in a study with 10th-grade students, Dimas et al. (2018) reported that some students could not create a correct table for data from an oscillator experiment. The students also had trouble connecting the data to the concept of harmonic motion (Dimas et al., 2018). Interpretation difficulties were reported in combination with all types of difficulties. They were described in combination with graph construction difficulties (Branisa & Jenisova, 2015; Jackson et al., 1993; Wu & Krajcik, 2006), variable ordering difficulties (Harrison et al., 2019; Nurrahmawati et al., 2021; Pratt, 1995), or both (Dewi et al., 2018; Onwu, 1993). Interpretation-related difficulties were also found in combination with data translation difficulties (Aydın-Güç et al., 2022; Branisa & Jenisova, 2015; Gültepe, 2016; Nurrahmawati et al., 2021). Several studies reported difficulties in combination with the STEM concept, e.g., graph construction (Berg & Phillips, 1994; Dimas et al., 2018), variable ordering (Karplus, 1979), or both (Brasell & Rowe, 1993). Furthermore, conceptual difficulties were found in combination with data translation difficulties (Oslington et al., 2020). Two studies identified student difficulties related to finding the correct graph type in combination with graph construction and variable ordering difficulties (Jackson et al., 1992; Ozmen et al., 2020). One study reported a part vs. whole difficulty in combination with graph construction difficulties (Åberg-Bengtsson, 2006).

## Discussion

This article presents a systematic review of empirical research on graphing in K–12 STEM education. A systematic search of three scientific databases found 12,945 records matching the search term. In total, we identified 50 relevant studies using our inclusion criteria, which were included in the review. Based on the codes presented above, we answer our research questions as follows. First, we summarize how graphing was implemented in the studies included in the literature review. Second, we condense the effectiveness of graphing as an instructional method, as reported in the studies. Third, we gather the reported student difficulties during graphing.

**Implementation of Graphing in K-12 STEM Education**

The theoretical foundation of graphing varied between studies: Many studies did not include a theoretical background, making an interpretation of the overarching results difficult because the study designs differed accordingly. A few studies included graphing from a cognitive perspective to construct a deeper understanding, e.g., by referring to the generation effect or the DeFT framework (Dimas et al., 2018; Rahmawati et al., 2020). In addition to the learning mechanisms during generative learning mentioned by Schmidgall et al. (2019), constructed graphs might offer additional cognitive functions by allowing learners to execute different cognitive strategies. For example, graphs can make inferences visible (Larkin & Simon, 1987). However, most studies investigate graphing owing to its relevance in professional or educational practices – that is, the aim is to learn graphing rather than using graphing as a learning method. Practices might vary between disciplines and between curricula, which in all likelihood played a role when investigating graphing in K–12 STEM education. Although most of the studies were conducted in the context of mathematics, a possible influence of the curriculum or the connection between graphing practices between disciplines could not be investigated in this review.

Regarding the specific aspects of graphing, line graphs were the most common type of graph constructed in our review. This could relate to the prevalence of bivariate data, which are often used to characterize relationships between two variables. Regarding the graphing method, there was no clear trend toward manual or tool-based graphing. However, a combination of tool-based and manual graphing seemed to facilitate graphing (Åberg-Bengtsson, 2006; Adams & Shrum, 1990; Ates & Stevens, 2003; English & Watson, 2015; Kramarski, 1999; Wu & Krajcik, 2006). Tool-based graphing was also used to analyze manually created graphs (Watson et al., 2022) or as an instructional method. For example, students who collected data via data acquisition probeware and used digital video analysis software in a physics course improved their graphing skills during the study (Struck & Yerrick, 2009).

Therefore, there was no typical study design that most of the studies used. However, an example of a common study design in this review would be an investigation of the effectiveness of graphing instruction for line graphs in the context of a regular class (during one or multiple lessons).

**Effectiveness of Graphing as an Instructional Method**

The second aim of this review was to outline the effectiveness of graphing as an instructional method, as described in the studies included in the literature review. This is challenging due to the differences in study design and methodology, e.g., tool-based vs. manual graphing and quantitative vs. qualitative analysis methods. Furthermore, several studies did not explicitly state a research hypothesis with clear results regarding the possible advantages of graphing.

Some studies reported that instruction helped students develop graphing skills (Gerard et al., 2012; Gultepe & Kilic, 2015; Meisadewi et al., 2017; Mevarech & Kramarsky, 1997; Struck & Yerrick, 2010; Wu & Krajcik, 2006). This is in line with Glazer (2011). However, the types of instruction varied between these studies and not all instructions referred specifically to graphing skills; other types of instructions, for example, in the context of specific topics (Struck & Yerrick, 2010; Wu & Krajcik, 2006) or scientific argumentation skills (Gultepe & Kilic, 2015), also facilitated students' graphing skills. This indicates that graphing skills can be improved using a broad range of instructions, provided that graphing is considered in some way during instruction.

Instruction was not only valuable to develop graphing skills but was also found to improve interpretation (Gerard et al., 2012; Gultepe & Kilic, 2015; Struck & Yerrick, 2010) and scientific process skills, such as developing hypotheses (Gultepe & Kilic, 2015).

Therefore, a focus on improving scientific process skills could be advantageous compared to traditional teaching (Gultepe & Kilic, 2015). Qualitative analyses reported similar results (Vitale et al., 2019; Wu & Krajcik, 2005). Due to the relevance of graphing for scientific inference (Gooding, 2010), this is an important aspect for teachers to consider during lesson planning.

Graphing skills seem to improve with grade level and age. In addition to Karplus (1979), six other studies considered this proposition (Garcia-Mila et al., 2014; Onwu, 1993; Padilla et al., 1986; Wavering, 1989; Webb & Boltt, 1991). This is related to mathematical understanding, which is also an important factor in graphing (Leinhardt et al., 1990). Two studies reported positive effects for manual graphing compared to tool-based graphing (Adams & Shrum, 1990; Branisa & Jenisova, 2015). Both studies compared manual graphing to graphing with (micro) computers. As there were 25 years between studies, and because technology evolved during this time, the benefit of manual graphing compared to letting a computer create a graph seems consistent, but the scant number of studies makes drawing conclusions difficult.

Several studies mention variables that might have a moderating effect on graphing skills, such as mathematical communication skills (Gardenia et al., 2021), statistical inference (Oslington et al., 2020), graph interpretation levels (Moritz, 2003), or cognitive development (Adams & Shrum, 1990), including mental structure (Berg & Phillips, 1994; Wavering, 1989) and scientific reasoning levels (Ates & Stevens, 2003). Unfortunately, due to the variance in the study designs and the low number of studies reporting moderators, a more detailed analysis of the effects of these moderators is not possible.

In conclusion, there seem to be multiple benefits of including graphing in K-12 STEM education (Glazer, 2011), such as improving graph interpretation skills. Graphing skills might also have the potential to facilitate scientific process skills. However, it should be noted that investigating the effects of graphing was not the main goal of most of the studies and there out of 50 included studies only nine had a control or comparison group. Most of the included studies investigated graphing in education from a practical perspective and focused on reporting the performance of their students.

### Difficulties During Graphing

Out of 50 included studies, 42 reported graphing difficulties. Difficulties during graph

construction were reported the most frequently. This included trouble with scaling (Åberg-

Bengtsson, 2006) or labeling the axes (Berg & Phillips, 1994). Variable ordering, such as sketching data points at the correct coordinates (Mevarech & Kramarsky, 1997), was also common. Furthermore, data translation, for example, translating data from a table to a graph, seemed to cause students problems (Tairab & Khalaf Al-Naqbi, 2004). These results highlight the relevance of metarepresentational competence (diSessa, 2004; Rau, 2017). This is also reflected in students' strategies, such as constructing appropriate data visualizations using self-questioning and reflecting (Chang et al., 2024).

Theoretical difficulties were not reported as often but were also found repeatedly. The

most common types of theoretical student difficulties were difficulties with interpretation

(n=12) and concept (n=11). For example, students chose the wrong graph type for the data or

STEM concept because they seemed to have problems connecting it to the context of the task and therefore could not construct a fitting graph (Jackson et al., 1992; Ozmen et al., 2020).

Similarly, students seemed to have trouble determining the x- and y-coordinates (Dewi et al.,

2018; Onwu, 1993), which could lead to "misunderstanding the graph" (p. 3, Dewi et al.,

2018). A possible reason for missing graph interpretation and construction skills could be missing practice (Tairab & Al-Naqbi, 2004).

In total, 20 studies reported difficulties in both categories (see Supplementary

Material). Therefore, theoretical difficulties might be related to construction difficulties. One study determined that "students with good levels of conceptual understanding were concluded to have strong graphing skills" (Gültepe, 2016, p. 53), whereas the opposite was found for students with low conceptual understanding. This connection between construction and theoretical difficulties is in line with Duval (2006) who considered translations within one register to be a possible cause of comprehension difficulties.

In summary, difficulties during the construction of a graph were consistently observed. Many of the included studies reported theoretical difficulties related to interpreting the data as well as convention-based difficulties during construction. This is in line with previous reviews that have also reported student difficulties; however, to our knowledge, none have specifically analyzed this connection (Clement, 1985; Leinhardt et al., 1990; Boels et al., 2019). These results indicate that students might have difficulty during graphing not only due to trouble understanding the conventions but also because they might not be able to correctly understand the data and its relevance and therefore might not know how to best display it.

### Implications for Practice

Based on our review results, we can join previous research (e.g., Glazer, 2011; Leinhardt et al., 1990) in emphasizing the relevance of graphing and encouraging teachers to add graphing activities to their lessons. The difficulties students experience during graphing exemplify the importance of graphing instruction. Graphing activities during the instruction of scientific argumentation (Gultepe & Kilic, 2015) might be an effective tool for improving science learning (Gerard et al., 2012). Similarly, comparing students who plotted the given data with those who plotted imagined data indicated that the first activity led to more integration between students' ideas and scientific evidence and, therefore, to a deeper exploration of the graphs (Vitale et al., 2019). This relates to the relevance of using authentic, real-world data that students can relate to when instructing data literacy (Friedrich et al., 2024). However, students analyzing the provided data graphed more accurately (Vitale et al., 2019). This suggests that students might benefit (1) from graphing their perception of scientific principles before traditional instruction about a topic and (2) from an in-depth analysis of the provided data.

## Limitations and Future Research

### Limitations

This research has several limitations. The most relevant one might be the large number of studies found in the initial search. Due to the large number of studies, we had to limit the included studies to those with a precise focus on graphing. However, there are many more studies in which students construct graphs—for example, as part of studies more broadly analyzing scientific inquiry skills—that might have provided valuable contributions to the topic. Furthermore, we included only peer-reviewed studies and no gray literature as quality control, and all included studies were written in English. As the publication years ranged from 1979 to 2024, these criteria may have changed and influenced our study selection. The choice of codes could have also led to limitations. The included studies often reported specific student difficulties. Due to the number of specific difficulties reported, we considered only overarching categories of difficulties in this review. The coding of these difficulties was often a part of the discussion between raters. Although raters always reached an agreement, a more fine-grained analysis might provide further insights.

### Future Research

Graphing as a method was often not based on theoretical research, but was justified due to its use in school (e.g., Moritz, 2003). We believe that more hypothesis-based testing grounded in theory in future research could provide valuable insights into the specific benefits of graphing in education. Additionally, more longitudinal research starting with younger students could generate a deeper understanding of the development of graphing skills. Further information about the influence of possible moderating variables could help improve instruction for students. Additionally, considering the function of graphing – whether students graph for themselves during learning, create a graph for others to explain something, or use a graph to compute a result during problem-solving – could play a role in interpreting the results. A future meta-analysis of empirical qualitative studies should take the effectiveness of graphing in these contexts into account.

## Conclusion

Understanding graphical representations of data is an important skill, and graphing is a relevant part of graph interpretation competence. Therefore, graphing has been examined in many studies. In this systematic literature review of 50 studies, we aimed to provide an overview of current research findings on graphing in K-12 education. We focus on the possible benefits of and difficulties faced by students during graphing, as well as how graphing is implemented in research studies. Studies have frequently analyzed graphing in the context of a course and have often found graphing instruction beneficial for not only improving graphing skills but also graph interpretation. However, the students experienced various graphing difficulties, such as correctly sketching data points and interpreting the graph. Therefore, the difficulties encountered during the construction of graphs might be related to the theoretical understanding of the data. Consequently, both types of difficulties should be considered during instruction, for example, first by graphing students' perceptions of a scientific phenomenon and then independently revising the graph based on the data measured during an experiment.

## List of Abbreviations

AERA: American Educational Research Association

CTML: Cognitive Theory of Multimedia Learning

DeFT framework: Design, Function, and Task framework

EARLI: European Association for Research on Learning and Instruction

ESERA: European Science Education Research Association

ICAP framework: Interactive, Constructive, Active, and Passive framework

ICLS: International Conference of the Learning Sciences

ICOTS: International Conference on Teaching Statistics

ITPC: Integrated Model of Text and Picture Comprehension

NARST: National Association for Research in Science Teaching

STEM: Science, Technology, Engineering, and Mathematics

PRISMA: Preferred Reporting Items for Systematic reviews and Meta-Analyses

## Declarations

### Availability of data and materials

All data generated or analysed during this study are included in this published article [and its supplementary information files].

### Competing interests

The authors declare they have no competing interests.

### Funding

The project on which this report is based was funded by the German Federal Ministry of Education and Research (BMBF) under the funding code 16MF1006B. The responsibility for the content of this publication lies with the authors.

**Authors' contributions**

Conceptualization, V.R., S.M, S.K., S.B., M.V., R.B., and J.K. ; methodology, V.R., S.M, S.K., S.B., M.V., R.B., and J.K.; formal analysis, V.R.; investigation, V.R., D.T., validation, V.R., D.T., S.M, S.K., S.B., M.V., R.B., and J.K; data curation, V.R., D.T.; writing—original draft preparation, V.R.; writing—review and editing, V.R., D.T., S.M, S.K., S.B., M.V., R.B., and J.K; visualization, V.R.; supervision, S.M., S.K., and J.K.; project administration, S.M., funding acquisition, J.K..